**A Finite Element Method for Simulation of Coupled Dynamics of Dislocations and Fracture**

Boyang Gu[1], Adrian Diaz[2], Yang Li[3] and Youping Chen[1]

[1] Department of Mechanical and Aerospace Engineering, University of Florida, Gainesville, FL 32611, USA
[2] Northrop Grumman Corporation, Linthicum, Maryland, USA
[3] Department of Physics and Astronomy, MSN 3F3, George Mason University, Fairfax, VA 22030, USA

## ABSTRACT

This work presents a finite element (FE) method for simulation of dynamic processes involving coupled dynamics of dislocation and fracture. The method is developed to numerically solve the Concurrent Atomistic–Continuum (CAC) representation of the conservation law of linear momentum. It discretizes a crystalline material at the unit cell level using 6-node prism elements, with the geometry of the elements permitting dislocations and cracks to nucleate and propagate along element boundaries. Nanoscale simulation results of single crystal Cu, Fe, and Si demonstrate the initiation and propagation of dislocations and cracks, as well as their interaction. These are reproduced by the FE method in excellent agreement with atomically resolved molecular dynamics simulations. Simulations of single crystal Cu at the mesoscale then demonstrate the efficacy of the FE method in capturing size-dependent brittle and ductile behavior. Specifically, a plane strain model of Cu is shown to fracture in a brittle manner, whereas with a three-dimensional model of Cu, curved and intersecting dislocations are observed to blunt the crack tip and prevent the propagation of the crack, leading to ductile behavior. The efficiency, accuracy, and applicability of the FE method are discussed.

## 1. INTRODUCTION

Dislocation slip, twinning, and cleavage are fundamental mechanisms of the deformation and failure of crystalline materials. They are discontinuous material behaviors and hence cannot be simulated using computational tools that are developed based on classical continuum mechanics. This is because the governing equations of classical continuum mechanics are represented by partial differential equations. Displacement discontinuities, such as those with dislocations and cracks, are thus not allowed in a classical continuum mechanics-based computer model, unless the slip planes or crack surfaces are modeled as surface boundaries. This is a well-known challenge in computational mechanics.

Many numerical methods have been developed to address this challenge by modifying the discretized governing equations to accommodate predefined or arbitrary discontinuities. Two representative methods are the cohesive zone model [1], which is based on a traction-displacement relation, and the extended finite element method (X-FEM) [2],which enriches the displacement approximation to incorporate discontinuous fields. A reformulation of elasticity, i.e., a nonlocal continuum mechanics theory called Peridynamics, has been developed by Silling [3], which replaces the spatial derivative of stress with internal force densities in the linear momentum balance equations and thus eliminates the need for continuity. However, all of these methods are developed based on a top-down continuum formulation and hence are challenging to model atomic-scale events such as the nucleation of dislocations or dislocation-crack interactions.

Computational tools such as Molecular Dynamics (MD) simulations have enabled our understanding of dynamic processes in nanoscale materials or phenomena, including the dynamics of dislocations [4-6] and the dynamics of fracture [7-9]. For mesoscale (100 nm to 100 μm) materials or phenomena, it is still challenging for state-of-the-art supercomputer MD to predict the dynamic processes and the underlying mechanisms [10].

There are several mesoscale or multiscale methods that can simulate the evolution of defects in space, e.g., dislocation dynamics (DD) [11], phase field method [12], and multiscale methods such as QC [13], BDM [14], CADD [15]. However, approaches such as DD, phase field, and CADD treat defect mobility using overdamped dynamics, and consequently the dynamics associated with moving defects are altered. QC requires near full atomistic refinement in dealing with the nucleation of defects such as dislocations [16] and has not been applied to solve dislocation or fracture problems beyond the nanoscale. The bridging-domain method [14] overlaps atomistic and continuum regions through an energy-based formulation. It is capable of fully dynamic simulations and has been applied to fracture problems. However, it is challenging for BDM to capture spontaneous nucleation or motion of dislocations, and the method has not been used for simulation of coupled dislocations and fracture or their dynamical interactions.

These limitations highlight the need for a multiscale method that can capture the nucleation and evolution of dislocations and cracks from the atomic to the mesoscopic or larger scales without the need of prescribed rules of criteria. The Concurrent Atomistic–Continuum (CAC) formulation was developed to address this need [17, 18]. Building on the Irving-Kirkwood (IK)'s statistical mechanical theory of transport processes [25], CAC extends the IK formulation from homogenized systems to polyatomic crystalline materials by incorporating the internal degrees of freedom of atoms within each unit cell into the field equations of conservation laws. This leads to a concurrent atomistic-continuum description of crystalline materials in which a crystal is represented as a continuously distributed lattice cell, with a discrete atomic basis being situated within each lattice cell, as that in solid state physics [19], as illustrated in Fig 1. The two-level structural description provides the physical foundation for a unified formulation that bridges atomistic and continuum descriptions of crystalline materials.

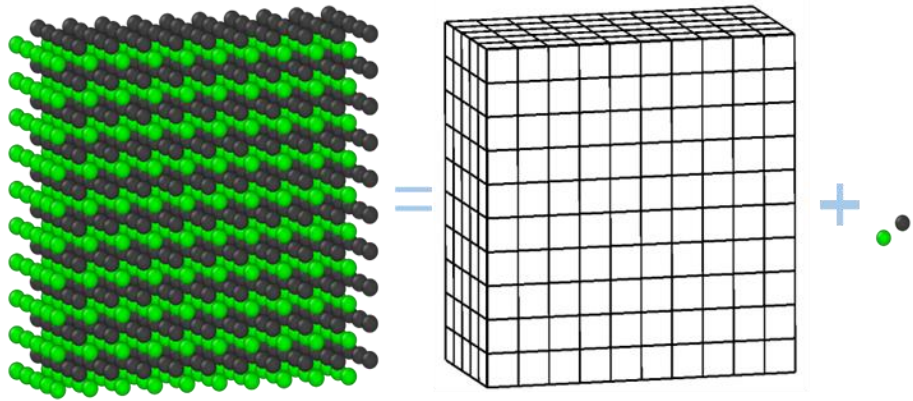

Fig. 1.  Crystal structure = lattice + basis.

Unlike domain-decomposition-based multiscale methods that consist of separate atomistic and continuum regions, CAC achieves concurrent atomistic and continuum description through its formulation of the governing laws and is therefore multiscale by formulation. Owing to its nonlocal formulation, CAC allows displacement discontinuity between finite elements and can thus naturally capture the nucleation, motion, and interaction of lattice defects such as dislocations and cracks through relative movements between neighboring elements.

Several parallel CAC codes based on a modified finite-element formulation have been developed, including PyCAC [20], CAC [21], and CAC-in-LAMMPS [22]. These codes resolve details to full atomic resolution at defect regions while using a finite element representation in ordered single crystalline regions. These CAC codes have successfully simulated dynamics of dislocations [23-32], grain boundaries [32-39], fracture [40-42], and defects associate with materials interfaces [30, 43, 44]. The CAC method has also simulated the coupled dynamics of defects, interfaces, and phonons [33, 45-47], including coherent and incoherent phonon scattering by dislocations [47] or grain boundaries [33] and phase interfaces [45, 46], as well as phonon-dislocation resonances, and their effects on dislocation dynamics and thermal transport [46].

Despite these capabilities, CAC implementations using eight-node hexahedral finite elements have a limitation in simulation of coupled slip and cleavage processes, as eight-node hexahedral elements cannot align their surfaces with both slip and cleavage planes if these planes do not coincide. To address this limitation, this work aims to introduce, test, and demonstrate a new discretization strategy.

The remainder of the paper is organized as follows. Section 2 briefly introduces the CAC formulation, numerical implementation, and the efficiency of the CAC finite element (FE) method. Section 3 presents a verification study of the method by comparing CAC simulation results of three representative crystal structures: face-centered cubic (FCC), body-centered cubic (BCC), and diamond-cubic crystals with that by molecular dynamics simulations. Section 4 presents mesoscale simulation results to demonstrate the capability of the method in capturing a size-dependent fracture phenomenon: the brittle fracture behavior of Cu under plane-strain conditions and the ductile fracture behavior of three-dimensional Cu crystals as a result of dislocation emissions from the crack tip. This paper ends in Section 5 with a brief summary and discussions.

## 2. Methodology

**2.1 CAC Governing Equations**

The Concurrent Atomistic-Continuum (CAC) formulation is an extension of Irving and Kirkwood's statistical mechanical theory for homogenized systems to a concurrently atomistic-continuum description of polyatomic crystalline materials. The formulation has evolved substantially over the past years. In particular, it has been recently reformulated using the theory of distributions for atomistic representation of fluxes[48-50] and the balance laws[51], and also for unifying the atomistic and continuum definitions of temperature[52]. In CAC, the local linear momentum density is defined as an average quantity over a time-step and a unit-cell volume as

$$\rho_\alpha \boldsymbol{v}_\alpha = \frac{1}{\Delta t} \int_t^{t+\Delta t} dt \frac{1}{V} \iiint_V d^3 r \sum_{k=1}^{N_l} m_\alpha \boldsymbol{v}_{k\alpha} \delta(\boldsymbol{r}-\boldsymbol{r}_{k\alpha}) \Box \sum_{k=1}^{N_l} m_\alpha \boldsymbol{v}_{k\alpha} \bar{\delta}_V(\boldsymbol{r}-\boldsymbol{r}_{k\alpha}) \quad (1)$$

where $\bar{\delta}_V$ is the average of the Dirac delta, $\delta$, over a time-step and unit cell volume. The distributional derivative of the linear momentum density has been obtained as [52]

$$\rho_\alpha \dot{\boldsymbol{v}}_\alpha = \boldsymbol{f}^{\text{int}}(\boldsymbol{r},t) - \frac{1}{\Delta t} \int_t^{t+\Delta t} dt \frac{1}{V} \oiint_{\partial V} \sum_{k=1}^{N_l} m_\alpha \tilde{\boldsymbol{v}}_{k\alpha} \tilde{\boldsymbol{v}}_{k\alpha} \delta(\boldsymbol{r}+\boldsymbol{r}'-\boldsymbol{r}_{k\alpha}) \cdot \boldsymbol{n} d^2 r' \Box \boldsymbol{f}^{\text{int}}(\boldsymbol{r},t) + \boldsymbol{f}^{\text{ext}}(\boldsymbol{r},t) + \boldsymbol{f}^T(\boldsymbol{r},t) \quad (2)$$

where $\boldsymbol{f}^{\text{int}}$ is the internal force density, $\tilde{v}_{k\alpha} \equiv v_{k\alpha} - v_\alpha$ is the difference between the particle velocity and the velocity field, and $\boldsymbol{f}^T(r,t)$ is related to thermal fluctuations or the kinetic part of stress, with its ensemble average being related to temperature [52]. At very low temperatures, such as the case studies used in this work, $f^T(r,t)$ is very small and hence can be neglected.

**2.2 Finite Element Discretization**

The CAC linear momentum equation can be solved using the finite-element method. To capture crystallographic defects such as dislocations and cracks, the geometry and orientation of the finite elements must reflect the crystal symmetry of the materials, and hence an appropriate element discretization strategy is required to align element surfaces with slip and cleavage planes.

The hexahedral elements employed in previous CAC implementations were constructed based on the parallelepiped-shaped primitive unit cells. This choice allows CAC to model slip or cleavage among the primary slip or cleavage planes, but not for dynamic processes that involve multiple slip planes and/or simultaneous slip and cleavage. To overcome this limitation, in this work we introduce a finite elements method based on conventional cubic or rectangular unit cells that have the full symmetry of the lattice, but each cube or rectangle further contains 8 prism elements. This discretization provides additional element surfaces that can be aligned with multiple slip and cleavage planes, thereby enabling the finite-element simulation of coupled dynamics of fracture and dislocations through the deformation and relative movements of the finite elements.

Fig. 2 illustrates the discretization, using a Si block as an example, by comparing three representations of the same model: (a) 8,000 atoms, (b) eight hexahedral elements, and (c) sixteen prism elements. It is noted that in single-crystal Si, dislocation glides occur predominantly on the {111} <110> slip systems, while cleavage may occur on both {111} and {110} planes. A comparison of Fig.2(b) and Fig.2(c) shows that the hexahedral elements expose only the {110} surfaces, whereas the prism elements introduce additional surfaces coincident with the {111} planes. The presence of these additional {111}-oriented surfaces enables the prism elements to represent both slip and cleavage in Si. Although Fig.2 is an illustration for Si, the same discretization principle can be applied to other crystal structures by adjusting element orientation to match their dominant slip and cleavage systems.

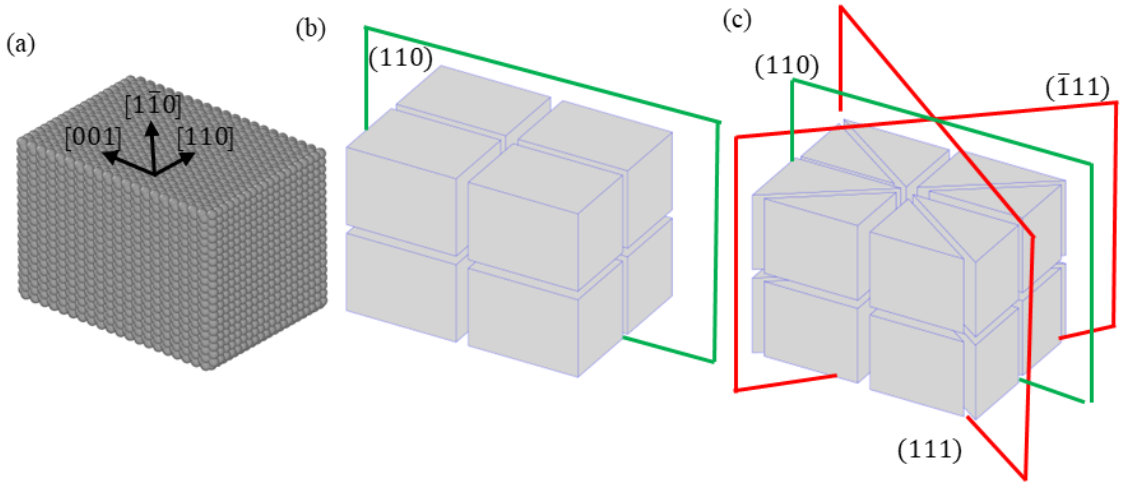

Fig. 2. A Si block is modelled by (a) 8,000 atoms, (b) 8 hexahedral elements, and (c) 16 prism elements. The {110} and {111} planes are indicated in green and red, respectively.

**2.3 Finite-Element Implementation with Prism Elements**

To numerically solve the governing equations in Eq. (2) using the finite element method, the displacement field within each element is interpolated using the shape functions:

$$\hat{u}_\alpha(x,t) = N_\xi(x) U_{\xi\alpha}(t) \tag{3}$$

where $U_{\xi\alpha}(t)$ is the $\alpha^{th}$ atom displacement at the $\xi^{th}$ node of the element and $N_\xi$ is the shape function at the $\xi^{th}$ node. Substituting Eq. (3) into Eq. (2), we obtain the weak form of Eq. (2) as

$$\iiint_{V_e} N_\xi(x) N_\eta(x) \rho_\alpha \ddot{U}_{\eta\alpha} dV = \iiint_{V_e} N_\eta(x) \left( f_\alpha^{int} + f_\alpha^{ext} + f_\alpha^T \right) dV \tag{4}$$

or in matrix form as

$$M_\alpha \ddot{U}_\alpha = F_\alpha \tag{5}$$

where $M_\alpha, U_\alpha, F_\alpha$ are the mass matrix, displacement vector and force vector,

$$M_\alpha = \iiint_{V_e} N_\xi(x) N_\eta(x) \rho_\alpha dV \tag{6}$$

$$F_\alpha = \iiint_{V_e} N_\eta(x) \left( f_\alpha^{int} + f_\alpha^{ext} + f_\alpha^T \right) dV \tag{7}$$

A schematic of the 6-node prism (wedge) in isoparametric space (s,t,w) is shown in Fig.3(a) and the shape functions are defined as

$$N_1 = (1-s-t)(1+w)/2, N_2 = s(1+w)/2, N_3 = t(1+w)/2,$$
$$N_4 = (1-s-t)(1-w)/2, N_5 = s(1-w)/2, N_6 = t(1-w)/2; \tag{8}$$

where $s \in [0,1], \quad t \in [0, 1-s], \quad w \in [-1,1]$.

Fig. 3(b) shows a distribution of Gaussian quadrature points for which the equivalent nodal force is computed via the weighted sum as follows

$$F_\alpha = F_{jk\alpha} = \sum_{i=0}^{n_q} w_i(s,t,w) f_j(s,t,w) N_k(s,t,w) \tag{9}$$

where $s,t,w$ are the isoparametric coordinates, $w_i$ is the weight of each quadrature point, and $n_q$ is the number of quadrature points within the element. $f_j$ and $N_k$ are the force and the shape function value at each quadrature point.

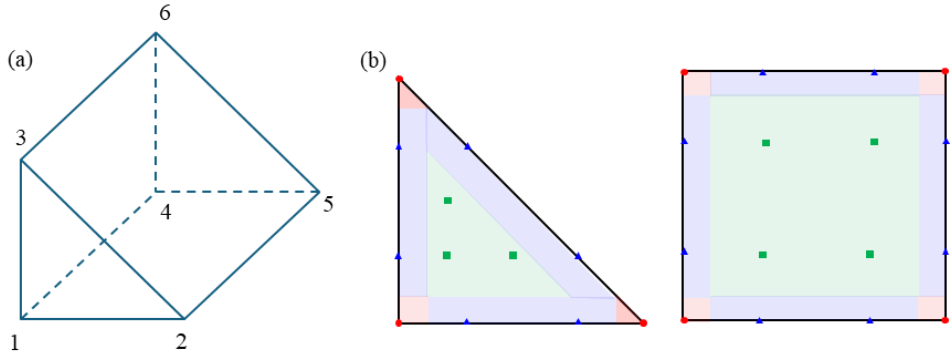

Fig. 3. (a) 6-node prism element in iso-parametric space with local axes and node numbering; (b) distribution of Gaussian quadrature points.

## 2.4 Computational Efficiency

Nanoscale simulations were conducted to quantify the computational efficiency of the CAC method with respect to molecular dynamics (MD). Fig. 4 compares the computational cost for simulation of single-crystal Cu blocks under dynamic relaxation. In the CAC simulations, the total number of prism elements was fixed at 2048, while the number of unit cells represented within each element was increased from 8×8×8 to 20×20×20. As a result, the computational cost of CAC remains nearly constant even as the physical size of the simulated domain increases. In contrast, the number of atoms in the MD models increases cubically with the system size, leading to a rapid increase in both memory demand and wall time.

As shown in Fig. 4, the ratio of memory usage per CPU (MD/CAC) increases with model size, rising from approximately 17 times for 10 unit cells per element length to 128 times for 20 unit cells per element length. A similar trend is observed for the MD/CAC ratio of total wall-time. For the largest model, the CAC simulation completed in 353 s, while the equivalent MD run required 4581 s, leading to a 13 times reduction in wall time. These results demonstrate that the finite element method significantly reduces the computational cost, in terms of memory and time, compared with MD.

(a)
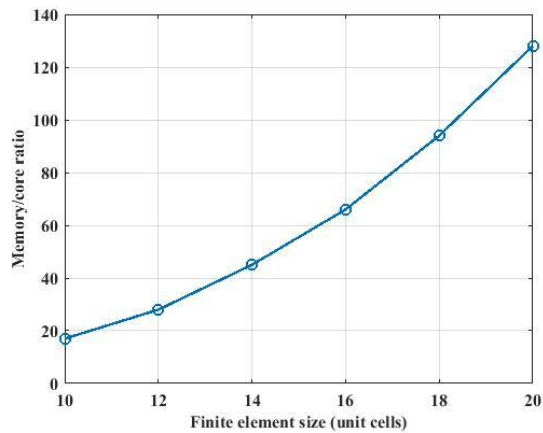

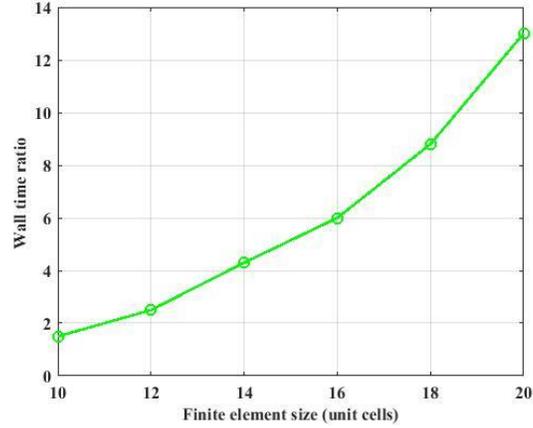

(b)

Fig. 4. Comparison of computational cost between CAC and MD for simulation of Cu blocks. (a) Ratio of memory usage per CPU (MD/CAC). (b) Ratio of total wall time (MD/CAC).

## 3. Verifications via Nanoscale Simulations

To test the accuracy of the finite element method, CAC simulations were performed to reproduce three types of phenomena: (i) dislocation emission from a crack tip in copper crystal under [110] tension, (ii) edge crack propagation in iron under [001] tension, and (iii) bending of a guided silicon beam.

### 3.1 Dislocation Emission from a Crack Tip in Copper

Copper was selected as a representative FCC crystal to test the finite element method. Two notched computer models with the same size and load conditions were constructed: (1) a CAC model containing 10,773 six-node prism elements (each pair of prism elements representing 1,728 atomic unit cells), and (2) a MD model containing 12,410,880 atoms. For both models, the simulation cell has dimensions of 104 nm × 120 nm × 12 nm, with crystallographic orientations aligned along [110], [001], and [1$\bar{1}$0], respectively. Periodic boundary conditions were imposed along the thickness direction ([$\bar{1}$10]). The interatomic interactions were described by the Cu–U3 EAM potential [53]. Tensile loading was applied along the [110] axis at a nominal strain rate of $7.69 \times 10^{-5}$ ps$^{-1}$.

In Fig. 5(a-b), we present a two-dimensional view of the finite-element deformation obtained from the CAC simulations at strains of 0.04% and 0.2%, respectively. Fig. 5(c–d) presents common neighbor analysis (CNA) of the simulation results, where green atoms represent the FCC lattice and red atoms denote stacking faults. Fig. 6(a-b) presents the finite element deformation and CNA results obtained from the MD simulation, respectively. As shown in Fig. 5, in each simulation, the pre-existing edge crack did not propagate during tensile loading. Instead, partial dislocations nucleated at the crack tip and glided away along two {111} slip planes. A comparison of Fig. 5 and 6 shows that the finite element method, using only 0.52% of the DOFs of the MD model, has reproduced dislocation emissions from the crack tip, in good agreement with the MD simulation.

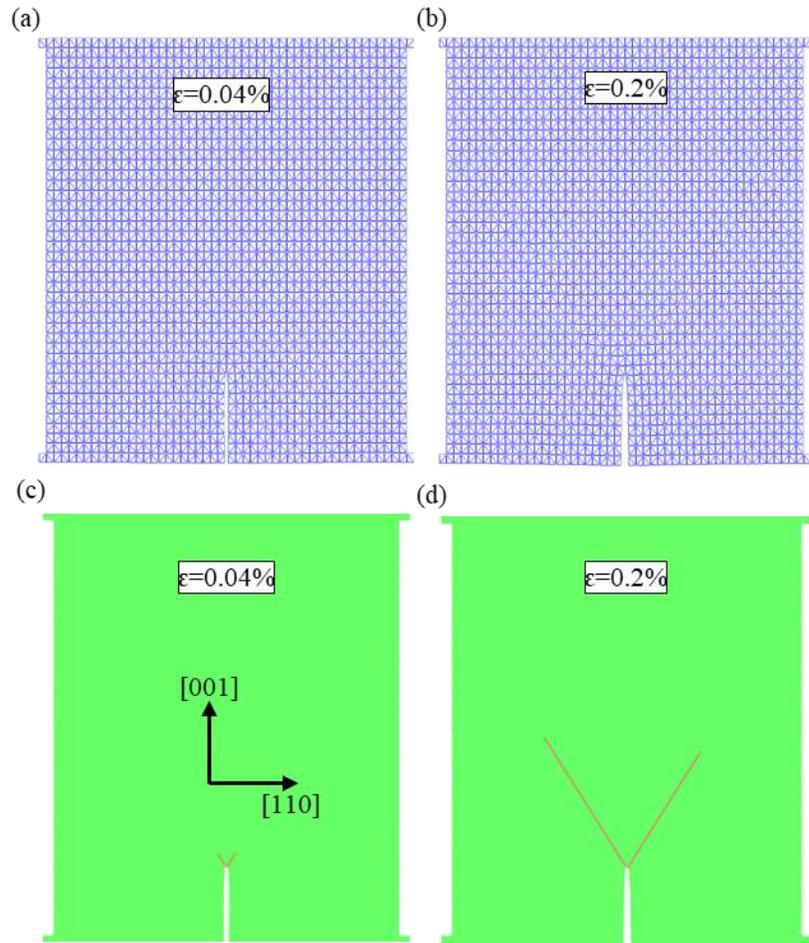

Fig. 5. Dislocation emission from a crack tip in Cu simulated by CAC with prism elements. (a–b) Finite-element deformation, and (c–d) the corresponding CNA results, where green indicates FCC atoms and red denotes stacking faults.

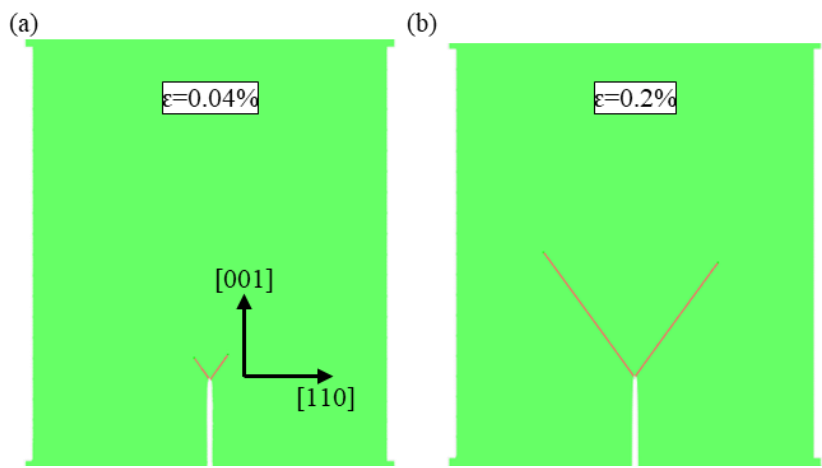

Fig. 6. The CNA results from MD simulations of Cu, where green indicates FCC atoms and red denotes stacking faults.

## 3.2 Crack Propagation in BCC Iron

Iron was selected as a representative BCC crystal to evaluate the finite element method. Three computer models were constructed: a CAC model discretized with 19,072 prism elements, a CAC model discretized with 9,536 hexahedral elements, and an atomically resolved MD model containing 2,384,000 atoms. The simulation cell has dimensions 100 nm×140 nm× 1.8nm. The crystallographic orientations were along [001], [110], and [$\bar{1}$10], respectively. A plane-strain condition was imposed in the out-of-plane direction ([$\bar{1}$10]) by constraining the displacement $u_z = 0$. An edge crack with length of 10 nm was introduced along the (001) cleavage plane. A displacement-controlled tensile loading was applied along the [001] axis at a nominal strain rate of $2 \times 10^8 \ s^{-1}$. Interatomic interactions were described by the Mendelev EAM-FS potential for Fe [54].

Fig. 7 presents a comparison of the simulation results among MD, CAC with prism elements, and CAC with hexahedral elements (top to bottom). As can be seen from Fig. 7, simulations of all three models have reproduced the propagation of the crack along the (001) plane. However, as shown in Fig. 7(c), only the CAC model with prism elements and the MD model have captured the nucleation of dislocations from the lower-right corner and their subsequent glide along the {112} slip planes. The CAC model with hexahedral elements, which lacks surfaces aligned with the {112} slip planes, failed to capture this dislocation nucleation and motion. Instead, it developed an unphysical separation of the computer model at the loading region. These results demonstrate the necessity of using prism elements for representing both slip and cleavage in Fe.

To provide a quantitative comparison of the three models, we extracted the strain at which the crack began to propagate and also the strain at which a dislocation nucleated. In addition, the crack propagation distance at 1.5% applied strain was measured. These results are summarized in Table 1, showing a close agreement between MD and CAC with prism elements in capturing the initiation and propagation of the crack and the dislocation.

Table 1. Summary of fracture and dislocation behavior in Fe.

| Model | Strain at which crack propagates | Strain at which dislocation nucleates | Crack propagation distance at 1.5% strain (nm) |
|---|---|---|---|
| CAC with prism elements | 0.47% | 0.82% | 34.2 |
| MD | 0.48% | 0.82% | 33.7 |
| CAC with hexahedral elements | 0.47% | N.A. | 32.2 |

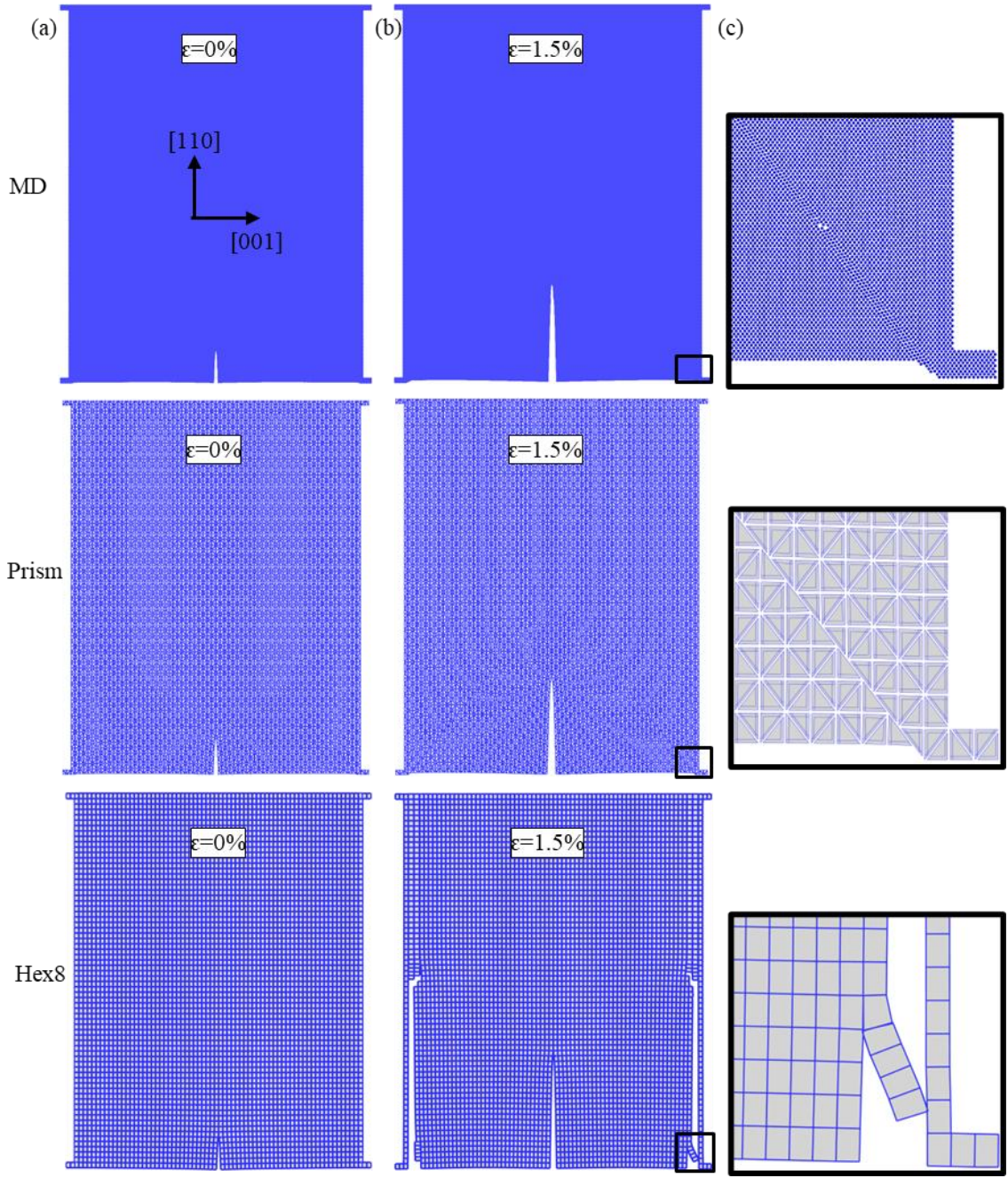

Fig. 7. Finite-element deformation of a pre-cracked Fe specimen under [001] tension. Columns: (a) initial crack (0% strain), (b) crack at ~1.5% strain, and (c) zoomed-in view showing dislocation nucleation on {112} slip planes. Rows correspond to MD, CAC with prism element and CAC with hexahedral element, showing only the MD and prism-CAC simulations capture dislocation emission from the lower-right corner.

### 3.3 Deformation of a Guided Silicon Beam

Analytical solution of the bending of a beam is widely used to test the accuracy of different finite elements. To quantify the accuracy of the finite element method in simulation of beam bending, we simulated a single-crystal Si beam of dimensions 19 nm × 130 nm × 19 nm. The displacements at the left end of the beam were fixed. The right end was subjected to a constant velocity $v = 0.1\ Å/ps$ in the [001] direction. The time step was 0.002 ps, which is equivalent to an incremental displacement of $\Delta u = 2 \times 10^{-4} Å$. Three computer models were constructed with the same loading and boundary conditions: a CAC model discretized with 10,016 prism elements, a CAC model discretized with 5,008 hexahedral elements, and an atomically resolved MD model that contains 1,252,000 atoms. Interatomic forces were described by the Pizzagalli-modified Stillinger–Weber potential [55]. To obtain a quasi-static configuration for comparison with the analytical solution, the loading displacement was held fixed at 2 nm and the entire specimen was subsequently relaxed for 50 ps to dissipate kinetic oscillations.

Fig. 8 presents the simulation result of the deformation of the guided beam at the beam displacement of 2 nm. Fig. 9 compares the deflection of the beam's top surface as a function of the position along the beam length. The simulation results are also compared with the analytical solution. As can be seen from Fig.9, all three models have reproduced the analytical cubic displacement field, indicating that MD, the CAC method with prism elements or hexahedral elements all can reproduce the nonlinear deformation.

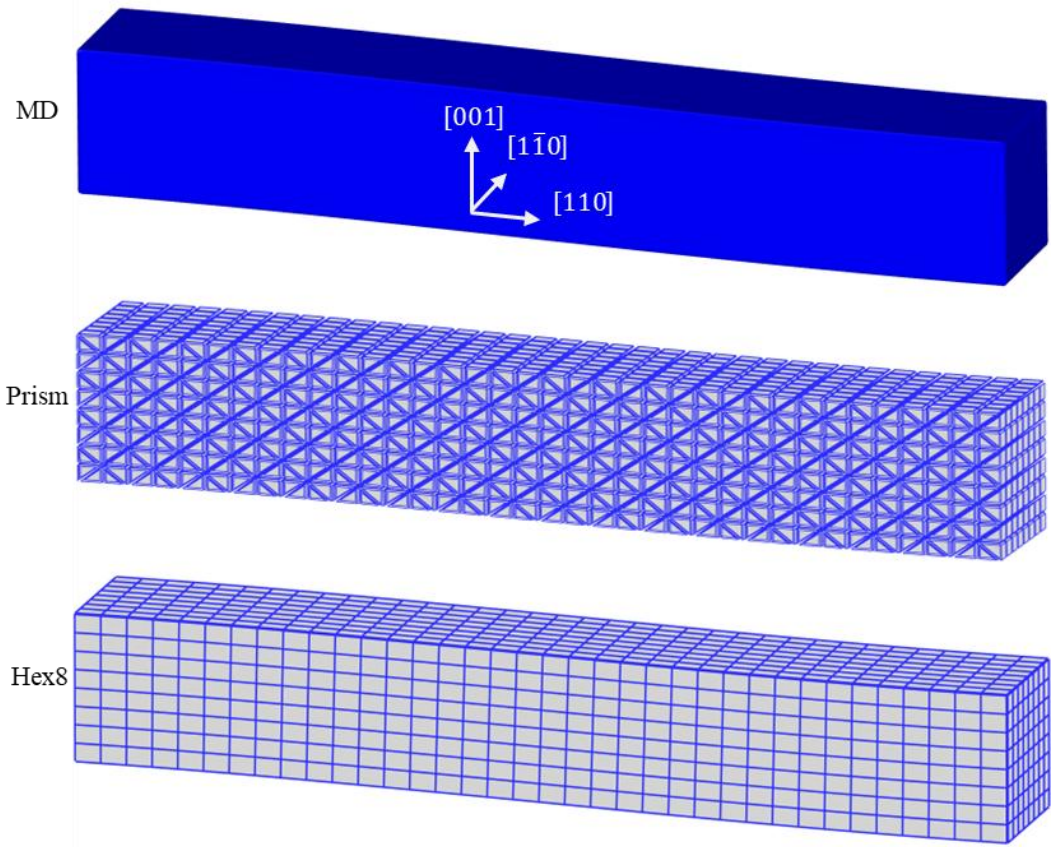

Fig. 8. A comparison of the deformation of a guided Si beam simulated by MD, CAC with prism elements, and CAC with hexahedral elements, respectively, under an end displacement of 2 nm.

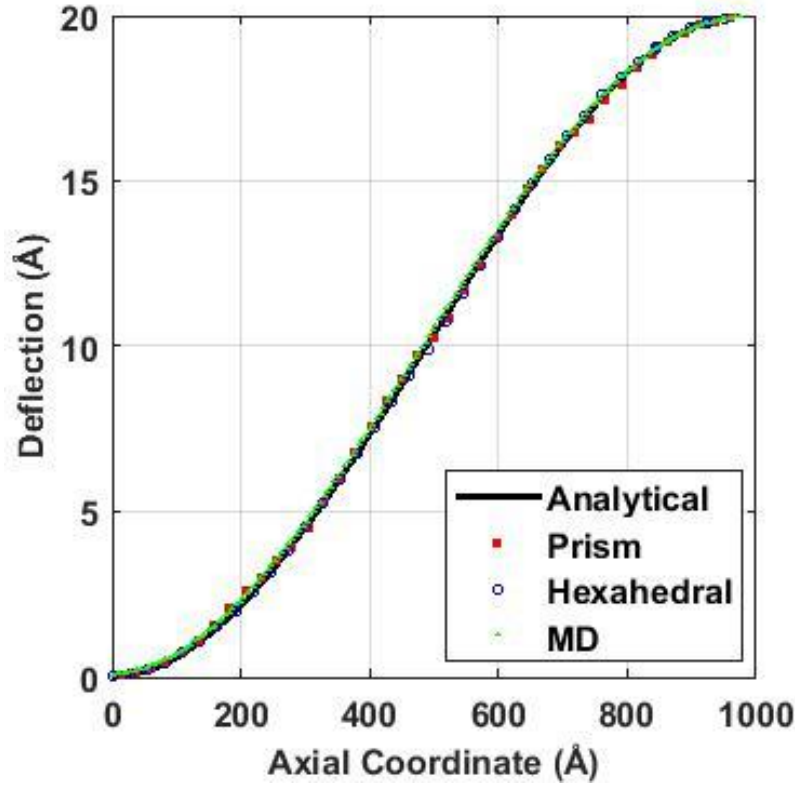

Fig. 9. Top surface deflection profiles from guided-beam bending of single-crystal Si.

### 4. Mesoscale simulations of size-dependent phenomena

Plane-strain models are commonly used in continuum mechanics as idealized two-dimensional descriptions of deformation, and they are widely employed in computational and experimental studies of fracture toughness.

To investigate the influence of plane strain on fracture behavior, we simulated a single-crystal copper specimen containing a pre-existing edge crack using the CAC method. The crystallographic axes were aligned with [001],[110], and [1$\bar{1}$0], respectively. Three models with identical in-plane dimensions of 300 nm × 300 nm but different thicknesses and boundary conditions were considered: (1) a 30 nm-thick model under a plane strain constraint, with out-of-plane displacement being fixed, (2) an unconstrained 30 nm-thick model, and (3) a fully three-dimensional (3D) model with thickness of 180 nm. The 30 nm- and 180 nm-thick models contain approximately $4.7 \times 10^5$ and $2.8 \times 10^6$ finite elements, corresponding to about $2.4 \times 10^8$ and $1.44 \times 10^9$ atoms in an equivalent atomically resolved MD model. The material interaction was described by the Mendelev EAM-FS potential for Cu [56], and tensile loading was applied to both the top and bottom surfaces along [001] at a strain rate of $1 \times 10^{-5}$ ps$^{-1}$.

Fig. 10 presents the finite-element deformation of the 30 nm-thick model under the plane-strain constraint. Simulation results show that the pre-existing crack propagates steadily along the {001} cleavage plane, exhibiting a brittle response. Fig. 11 presents a zoomed-in front-view of CNA visualization of dislocations and stacking faults near the crack-tip region. Partial dislocations are found to emit from the crack tip, activating multiple {111} slip systems as the crack advances. Despite the emission of dislocations from the crack tip, the crack continuously propagates along the {001} direction.

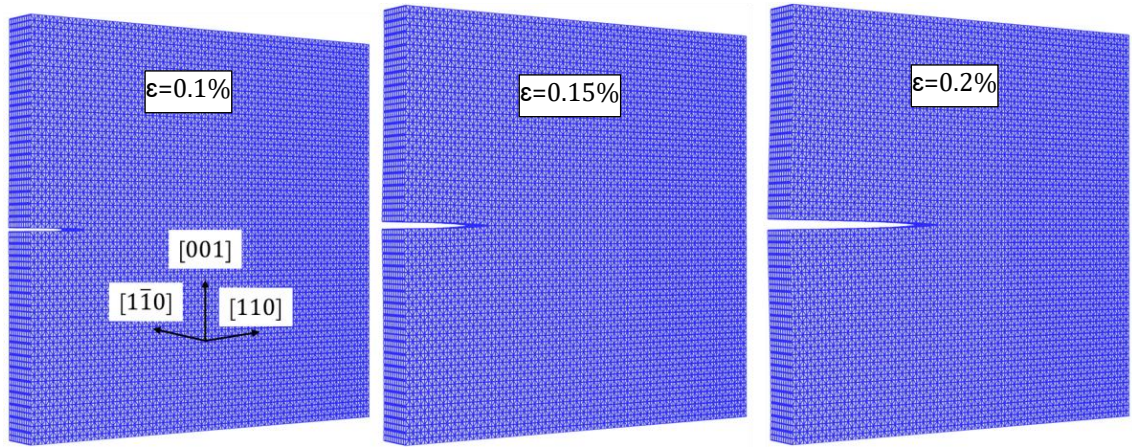

Fig. 10. Finite-element deformation of the 30 nm-thick Cu model under plane-strain constraint at various applied strains.

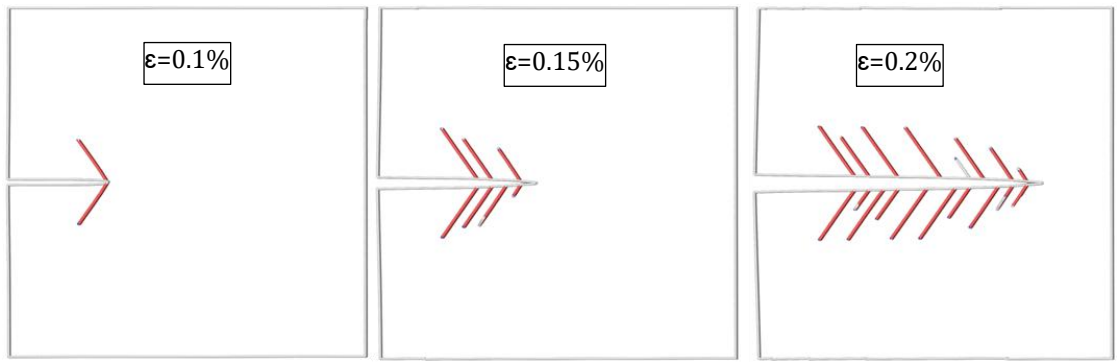

Fig. 11. Front-view CNA visualization of dislocations and stacking faults in the zoomed-in crack-tip region of the 30 nm-thick Cu model under plane-strain constraint at various applied strains.

Fig. 12 presents the deformaiton of the finite elements in the 30 nm-thick model without the plane-strain constraint. In contrast to the plane-strain case, the pre-existing edge crack along [110] does not propagate. Rather, the crack tip becomes blunted. Fig. 13 presents a zoomed-in front-view of CNA visualization of the crack-tip region, showing the evolution of dislocations and stacking faults. The emitted partial dislocations curve as they glide on the two active {111} slip planes adjacent to the crack tip, generating extended stacking faults that progressively blunt the crack tip. The absence of the out-of-plane constraint therefore promotes plastic relaxation and suppresses cleavage propagation,leading to a ductile response compared with the brittle behavior observed under plane-strain constraint.

Fig. 14 presents the finite-element deformation near the crack tip in the 180 nm-thick model. The deformation behavior is similar to that of the unconstrained 30 nm-thick model, where the crack is blunted as dislocations are emitted from the crack tip.

Fig. 15 presents the zoomed-in view of CNA visualization of dislocation and stacking-fault structures in the fully 3D models with 30 nm and 180 nm thickness, respectively. In the 30 nm model, dislocations move independently, exhibiting limited interaction on the same slip plane. In contrast, the 180 nm model develops more curved dislocation lines that extend over larger regions, resulting in intersections and short connecting segments between neighboring dislocations on the same {111} slip plane. Increasing the model thickness promotes the formation of complex three-dimensional dislocation networks, which is characteristic of ductile fracture.

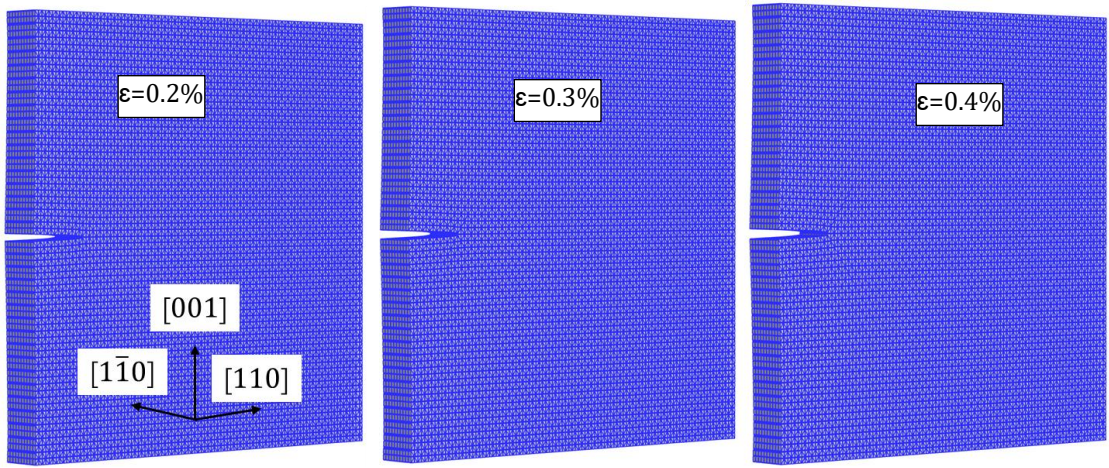

Fig. 12. Finite-element deformation of the 30 nm-thick Cu model, without the plane-strain constraint, at various applied strains.

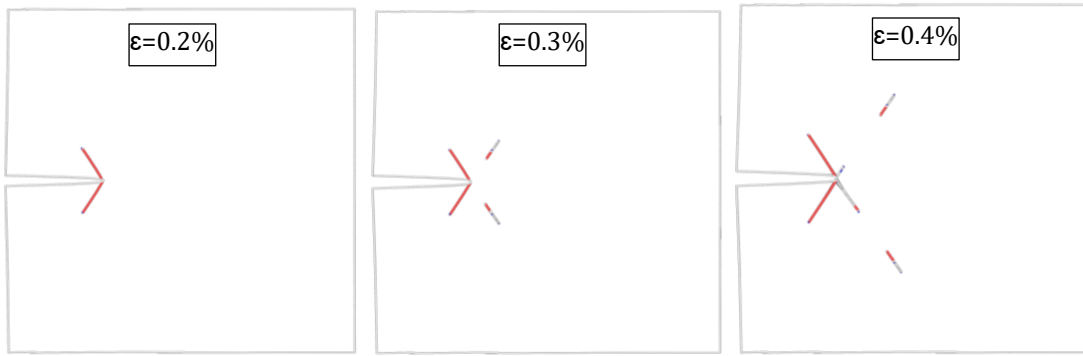

Fig. 13. Front-view of CNA visualization of dislocations and stacking faults in the zoomed-in crack-tip region of the 30 nm-thick Cu model at various applied strains.

In summary, the plane-strain constraint significantly affects the fracture dynamics that involve dislocations. Although multiple {111} slip systems are activated during crack advancement, the emitted dislocation partials remain short and nearly straight, resulting in a brittle fracture behavior. Removing the constraint allows curved dislocation partials on the active {111} slip planes to blunt the crack tip, leading to a ductile behavior. With a three-dimensional finite size computer model, increasing the model thickness to 180 nm provides the space for the longer dislocation lines and their interactions, including formation of short intersection segments that were absent in the 30 nm-thickness model. These results highlight the importance of fully 3D modeling for capturing the competition between crack advance and plastic relaxation.

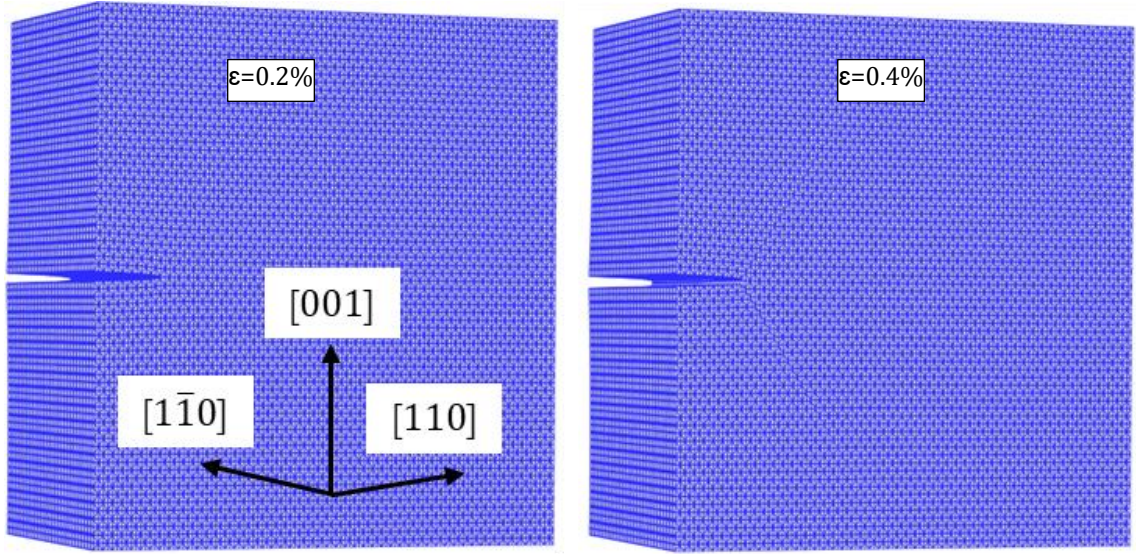

Fig. 14. Finite-element deformation of the 180 nm-thick Cu model at various applied strains.

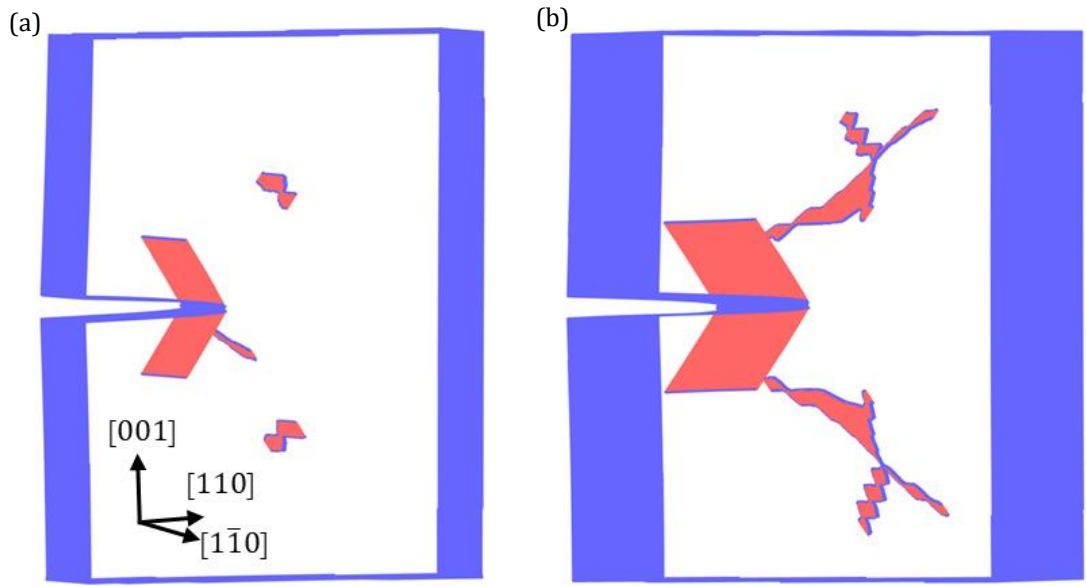

Fig. 15. CNA visualization of dislocation and stacking-fault structures in the zoomed-in crack-tip region (70 nm× 100 nm) for two finite-sized 3D models: (a) a 30 nm-thick model, and (b) an 180 nm-thick model, both at an applied strain of 0.3%.

## 5. Summary

In this work, we have presented a finite element method that can simulate spontaneous nucleation, glide, and interaction of dislocations, as well as crack propagation, and dislocation emissions from crack tips. The method builds on the CAC formulation and extends the CAC methodology by employing a six-node prism element. The finite element method allows both dislocation slip and cleavage to naturally emerge in a simulation through the deformation and relative movements of finite elements, as the direct consequence of Newton's second law and an interatomic potential, without the need for any empirical rules and parameters or special numerical treatments.

The efficiency and accuracy of the FE method was quantified through comparing simulation results obtained using the FE method with those from atomically resolved MD simulation for three nanoscale material phenomena: dislocation emission from a crack tip in Cu, crack propagation in Fe, and the bending of a guided Si beam. These simulations have shown that the FE method reproduces the coupled dynamics of dislocations and fractures as accurately as MD, but with significantly less memory usage and simulation time, indicating the capability of the method for larger-scale simulation than that can be achieved by MD.

A set of mesoscale simulations of notched single-crystal Cu was then performed. FE simulation results show that the plane-strain constraint significantly influences the local deformation at the crack tip, suppressing out-of-plane dislocation activities, and consequently promoting crack advance in a brittle manner. In contrast, FE simulations of a fully three-dimensional model capture the nucleation and propagation of curved dislocations that blunt the crack tip and impede crack propagation. Moreover, increasing the model thickness enables the development of extended and intersecting dislocation networks characteristic of realistic three-dimensional plasticity. These results demonstrate the effect of size on brittle and ductile fracture behaviors. They also indicate that large-scale three-dimensional simulations are necessary for understanding and predicting ductile fracture in real crystalline materials.

**Acknowledgement**

This work is based on research supported by the US National Science Foundation under Award Number CMMI-2054607. The work of YC is also partially supported by the US National Science Foundation under Award Number CMMI-2349160. The computer simulations are funded by the Advanced Cyberinfrastructure Coordination Ecosystem: Services & Support (ACCESS) allocation TG-DMR190008.